\documentclass[3p,times]{elsarticle}

\usepackage{ecrc}

\volume{00}

\firstpage{1}

\journalname{Nuclear Physics A}

\runauth{}

\jid{npa}

\jnltitlelogo{Nuclear Physics A}

\usepackage{amssymb}

\biboptions{square,comma,numbers,sort&compress}

\usepackage[figuresright]{rotating}

\begin{document}

\begin{frontmatter}



\dochead{}

\title{Dynamical Evolution, Hadronization and Angular De-correlation of Heavy Flavor in a Hot and Dense QCD Medium}


\author[duke]{Shanshan Cao}
\address[duke]{Department of Physics, Duke University, Durham, NC 27708, USA}
\author[ccnu]{Guang-You Qin}
\address[ccnu]{Institute of Particle Physics and Key Laboratory of Quark and Lepton Physics (MOE), Central China Normal University, Wuhan, 430079, China}
\author[duke]{Steffen A Bass}

\begin{abstract}
We study heavy flavor evolution and hadronization in relativistic heavy-ion collisions. The in-medium evolution of heavy quarks is described using our modified Langevin framework that incorporates both collisional and radiative energy loss mechanisms. The subsequent hadronization process for heavy quarks is calculated with a fragmentation plus recombination model. We find significant contribution from gluon radiation to heavy quark energy loss at high $p_\textnormal{\tiny T}$; the recombination mechanism can greatly enhance $D$ meson production at medium $p_\textnormal{\tiny T}$. Our calculation provides a good description of $D$ meson nuclear modification at the LHC. In addition, we explore the angular correlation functions of heavy flavor pairs which may provide us a potential candidate for distinguishing different energy loss mechanisms of heavy quarks inside QGP.  
\end{abstract}


\end{frontmatter}



\section{Introduction}

Heavy quarks serve as excellent probes of the highly excited and color deconfined QCD matter -- known as the Quark-Gluon Plasma (QGP) -- created in ultrarelativistic heavy-ion collisions. They are mainly produced at the primordial stage of the collisions and then propagate through and probe the whole evolution history of the QGP fireballs. Because of their large masses, heavy quarks were usually expected to interact more weakly with the medium than light partons. However, experimental data from at both RHIC and the LHC have revealed surprisingly small values of $R_\textnormal{\tiny AA}$ and large values of $v_2$ for heavy flavor mesons and their decay electrons \cite{Adare:2010de,Tlusty:2012ix,Grelli:2012yv,Caffarri:2012wz}. Various theoretical frameworks have been constructed to explore this ``heavy flavor puzzle", such as the Boltzmann-based parton cascade model  \cite{Uphoff:2012gb}, the linearized Boltzmann transport of heavy quarks inside a hydrodynamic medium \cite{Nahrgang:2013saa}, and the Langevin evolution of heavy quarks inside QGP \cite{Cao:2013ita}.

We follow our previous work \cite{Cao:2011et,Cao:2012jt,Cao:2013ita,Cao:2013wka} and study the dynamical evolution and hadronization of heavy quark in heavy-ion collisions. The energy loss of heavy quarks in QGP medium is described using a modified Langevin equation that simultaneously incorporates mechanisms of quasi-elastic scattering and medium-induced gluon radiation. The subsequent hadronization process is calculated with a hybrid fragmentation plus recombination model. Within this new framework, we demonstrate that while the collisional energy loss dominates heavy flavor nuclear modification at low $p_\textnormal{\tiny T}$ region, gluon radiation dominates at high $p_\textnormal{\tiny T}$. The recombination mechanism may significantly enhance the production of heavy mesons at medium $p_\textnormal{\tiny T}$. Our calculation provides a good description of $D$ meson $R_\textnormal{\tiny AA}$ as measured by the ALICE collaboration. We also study the angular correlation functions of heavy flavor pairs and find them sensitive to different energy loss mechanisms. 

\begin{figure}[tb]
\begin{minipage}{0.47\textwidth}
\includegraphics[width=0.96\textwidth,clip=]{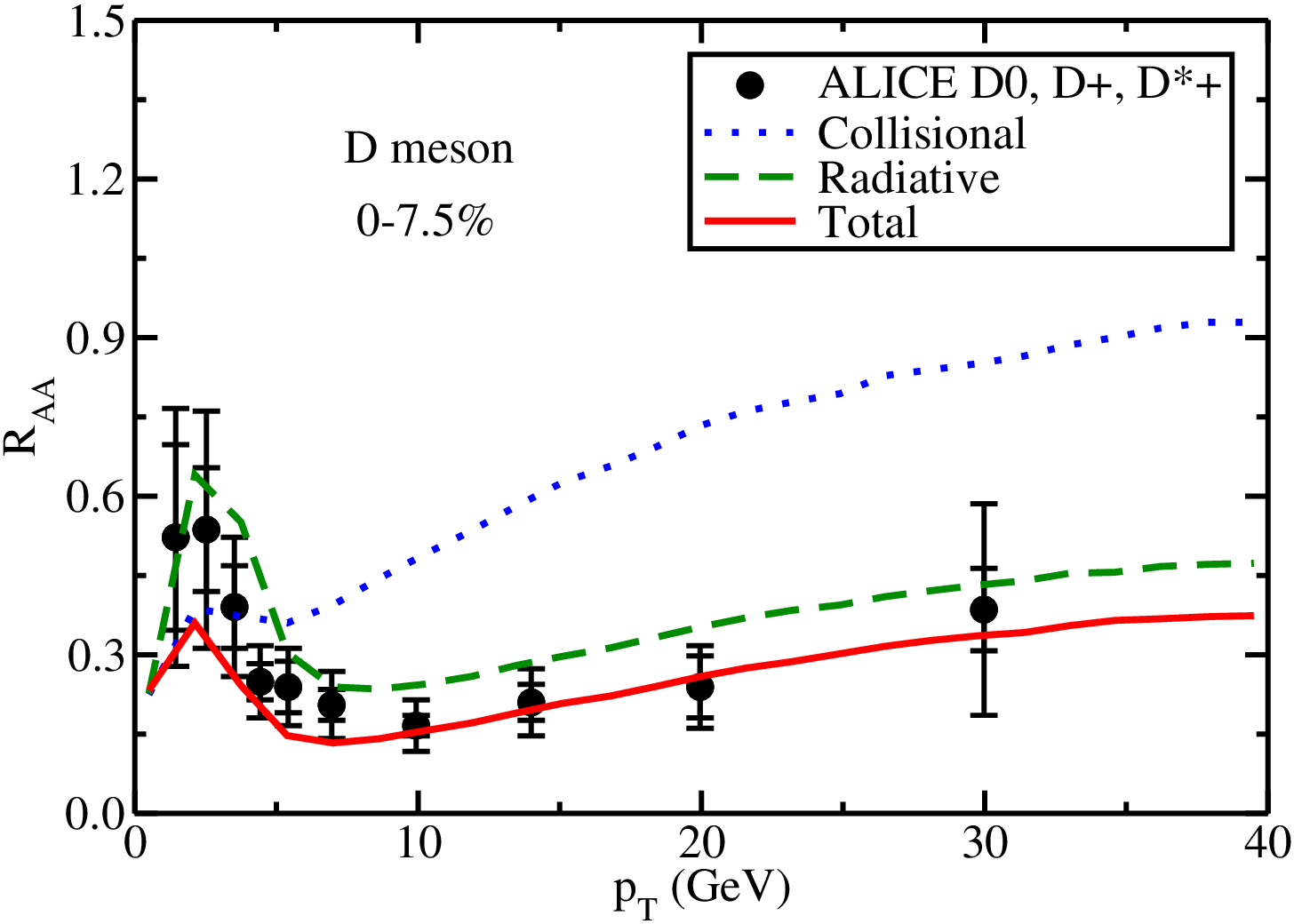}
\caption{\label{fig:RAA_DD6_compare_3mechanisms}(Color online) Comparison of $D$ meson $R_\textnormal{\tiny AA}$ between different energy loss mechanisms.}
\end{minipage}\hspace{0.06\textwidth}
\begin{minipage}{0.47\textwidth}
\includegraphics[width=0.96\textwidth,clip=]{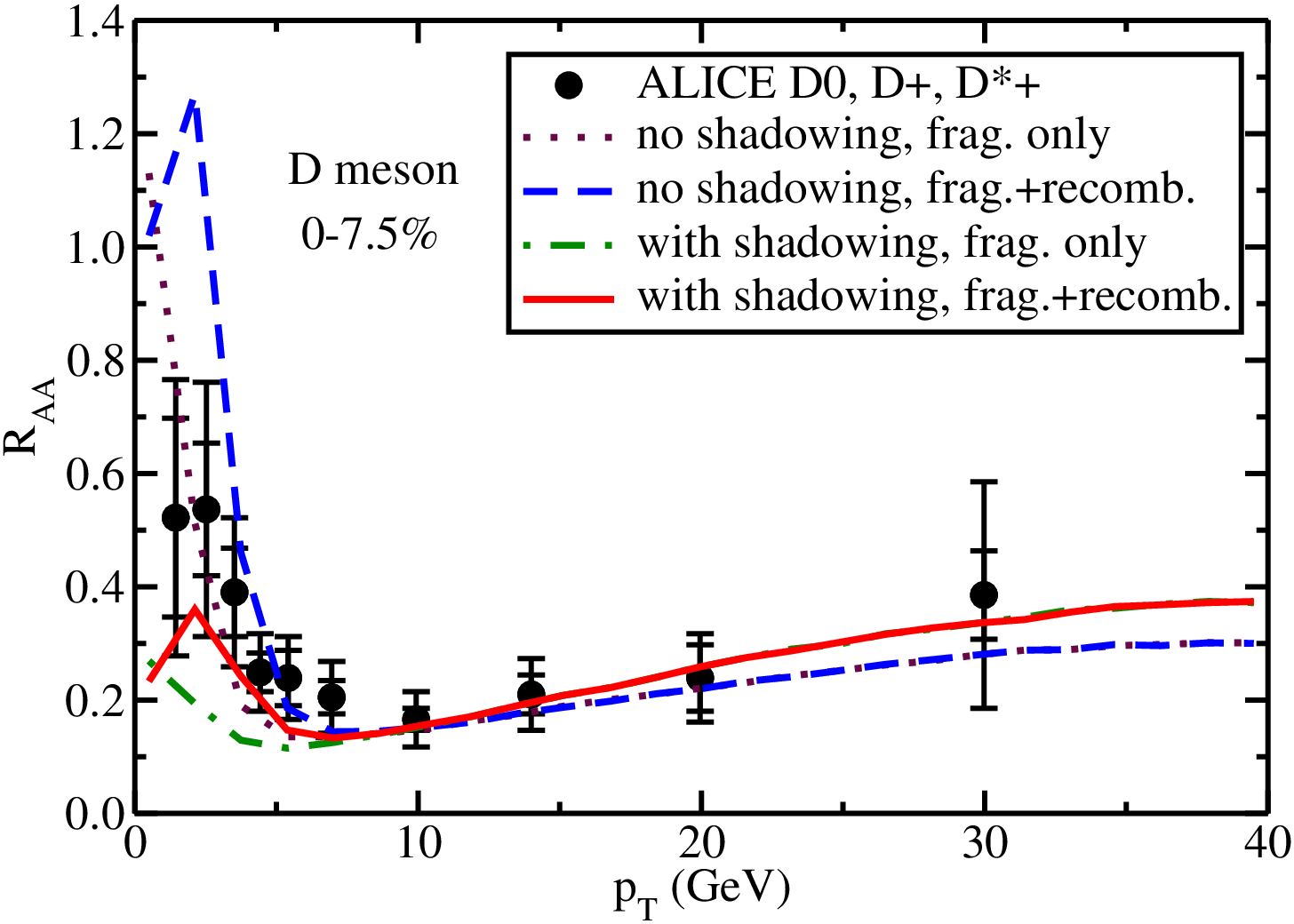}
\caption{\label{fig:RAA_DD6_compare_initial_final}(Color online) Comparison between different initialization and hadronization methods. This figure is taken from Ref. \cite{Cao:2013ita}}
\end{minipage}
\end{figure}

\section{In-medium Evolution and Hadronization of Heavy Quarks}

In the limit of multiple scatterings, the dynamical evolution of heavy quarks inside QGP medium can be treated as the Brownian motion that is typically described by the Langevin equation. To incorporate the effects from both quasi-elastic scatterings and medium-induced gluon radiation, we modify the classical Langevin equation as follows:
\begin{equation}
\frac{d\vec{p}}{dt}=-\eta_D(p)\vec{p}+\vec{\xi}+\vec{f_g}.
\end{equation}
Apart from the first two terms on the right hand side that represent the drag and the thermal random forces, a recoil force term $\vec{f_g}=-d\vec{p_g}/dt$ is introduced to describe the effect of gluon radiation on heavy quark motion. The probability of gluon radiation during each time interval $\Delta t$ and the momentum of radiated gluon ($\Delta\vec{p_g}$) are simulated using the Monte-Carlo method according to the following equation, taken from the Higher-Twist energy loss formalism \cite{Zhang:2003wk}:
\begin{equation}
\label{Wang}
\frac{dN_g}{dx dk_\perp^2 dt}=\frac{2\alpha_s(k_\perp)}{\pi} P(x) \frac{\hat{q}}{k_\perp^4} \textnormal{sin}^2\left(\frac{t-t_i}{2\tau_f}\right)\left(\frac{k_\perp^2}{k_\perp^2+x^2 M^2}\right)^4,
\end{equation}
where $\hat{q}$ is the gluon transport coefficient, $k_\perp$ is the gluon transverse momentum, $x$ is the fractional energy carried by radiated gluon, $\tau_f$ is the gluon formation time, and $P(x)$ is the splitting function. By requiring that heavy quarks approach thermal equilibrium after a sufficiently long time, one obtains the fluctuation-dissipation relation between the drag and the thermal force -- $\eta_D(p)=\kappa/(2TE)$, where $\kappa$ is the momentum space diffusion coefficient defined in $\langle\xi^i(t)\xi^j(t')\rangle=\kappa\delta^{ij}\delta(t-t')$. 
For radiative process, we set a lower cutoff $\omega_0=\pi T$ for gluon energy to incorporate the balance between emission and absorption. Different transport coefficients are related via $D=2T^2/\kappa$ and $\hat{q}=2\kappa C_A/C_F$, where $D$ is the spatial diffusion coefficient of the heavy quark in the QGP. For the results presented in this section, the diffusion coefficient is set as $D=6/(2\pi T)$.

We use our modified Langevin framework to simulate the heavy quark evolution inside QGP. The initial momentum distributions of heavy quarks are calculated using the leading-order pQCD approach unless otherwise specified. To calculate the partonic cross sections, we adopt CTEQ parton distribution functions \cite{Lai:1999wy} and include the nuclear shadowing effect using the EPS08 parametrization \cite{Eskola:2008ca}. After heavy quarks traverse the medium, their hadronization is simulated according to a hybrid fragmentation plus coalescence model developed in Ref. \cite{Cao:2013ita}.  In our work, the QGP is simulated with a (2+1)-dimensional viscous hydrodynamic model (VISH2+1) \cite{Song:2007fn,Song:2007ux}. We employ the code version and parameter tunings for 2.76~TeV PbPb collisions at the LHC that were previously used in Ref. \cite{Qiu:2011hf}.

In Fig.\ref{fig:RAA_DD6_compare_3mechanisms}, we compare $D$ meson $R_\textnormal{\tiny AA}$ between different energy loss mechanisms. One observes that the collisional energy loss dominates the low $p_\textnormal{\tiny T}$ regime while gluon radiation dominates high $p_\textnormal{\tiny T}$. Our combination of the two mechanisms provides a good description of the experimental data from the ALICE collaboration. In Fig.\ref{fig:RAA_DD6_compare_initial_final}, we investigate the effects of the nuclear shadowing effect in the initial production and the recombination mechanism in the hadronization process on heavy flavor quenching and find that the shadowing effect significantly suppress $D$ meson $R_\textnormal{\tiny AA}$ at low $p_\textnormal{\tiny T}$ while the heavy-light quark recombination enhances the production rate of $D$ meson at medium $p_\textnormal{\tiny T}$. More numerical results, such as the heavy meson $v_2$ and comparisons with the RHIC data can be found in Ref. \cite{Cao:2013ita}.

\section{Angular Correlation Functions of Heavy Flavor Pairs}

\begin{figure}[tb]
\begin{minipage}{0.47\textwidth}
\includegraphics[width=0.96\textwidth,clip=]{RAA_fit_with_each_mechanism.eps}
\caption{\label{fig:RAA_fit_with_each_mechanism}(Color online) Fitting $D$ meson $R_\mathrm{AA}$ with different energy loss mechanisms by tuning the diffusion coefficient.}
\end{minipage}\hspace{0.06\textwidth}
\begin{minipage}{0.47\textwidth}
\includegraphics[width=0.96\textwidth,clip=]{corr_compare.eps}
\caption{\label{fig:corr_compare}(Color online) Comparison of the angular correlation functions of $c\bar{c}$ pairs  between different energy loss mechanisms.}
\end{minipage}
\end{figure}

\begin{figure}[tb]
\begin{minipage}{0.47\textwidth}
\includegraphics[width=1.05\textwidth,clip=]{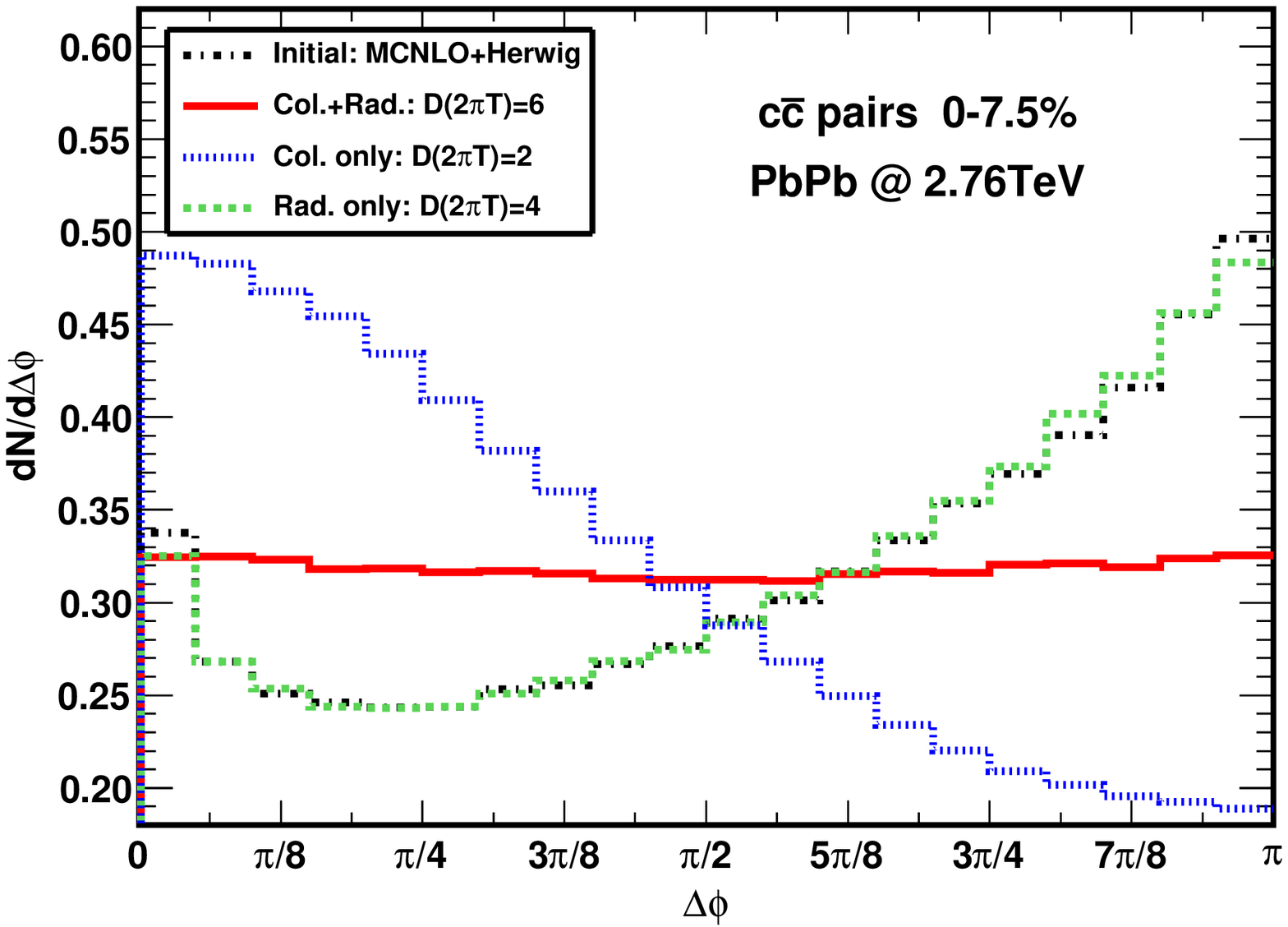}
\caption{\label{fig:ccbar-show}(Color online) Angular correlation functions of $c\bar{c}$ pairs with MCNLO+Herwig initial conditions.}
\end{minipage}\hspace{0.06\textwidth}
\begin{minipage}{0.47\textwidth}
\includegraphics[width=1.05\textwidth,clip=]{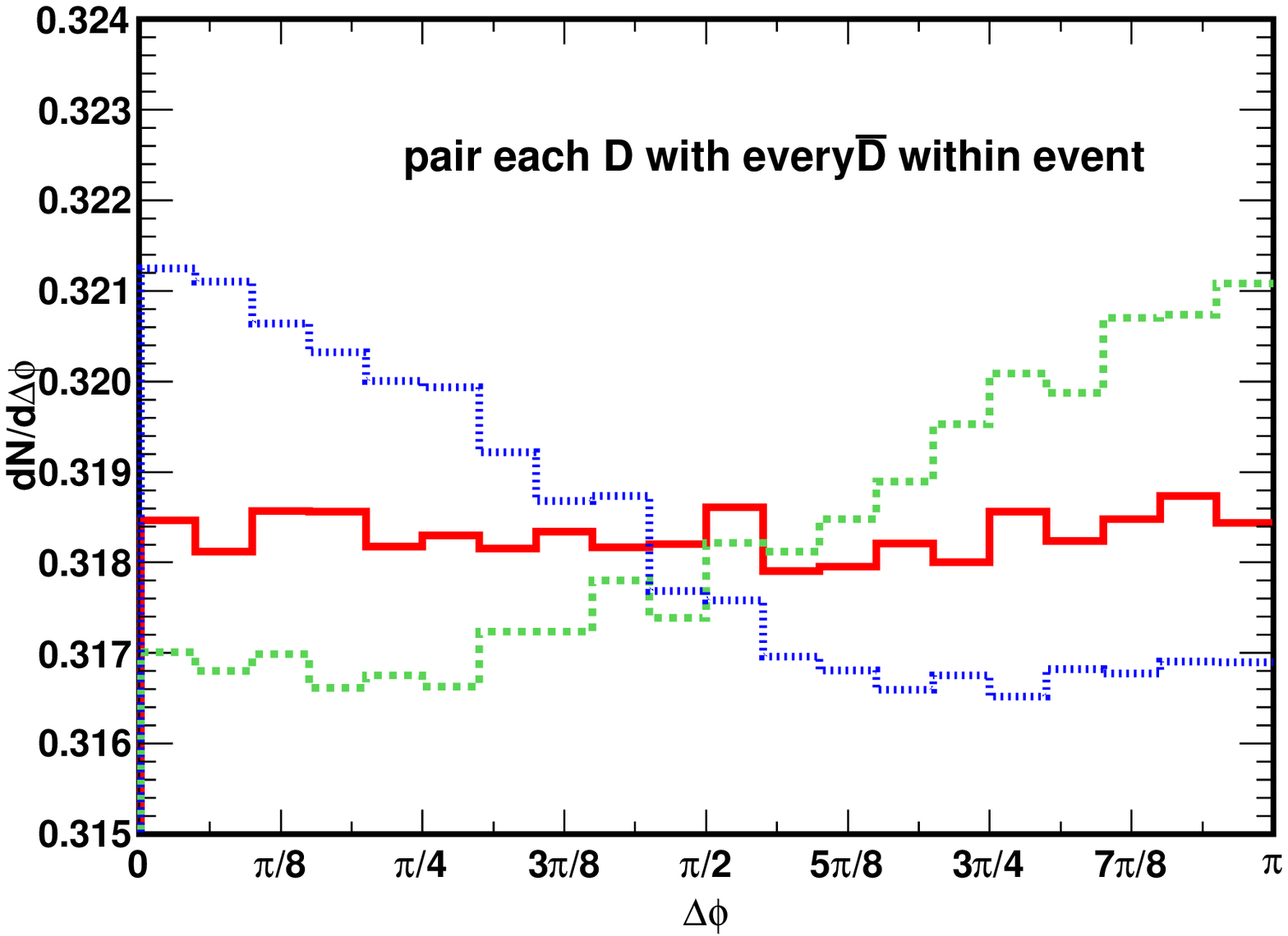}
\caption{\label{fig:DDbar-show}(Color online) Angular correlation functions of $D$-$\bar{D}$ with MCNLO+Herwig initial conditions.}
\end{minipage}
\end{figure}

In the previous section, we have shown a good description of $D$ meson $R_\textnormal{\tiny AA}$ when incorporating both collisional and radiative energy losses for heavy quarks. One may wonder how each energy loss mechanisms alone can describe the data; this is shown in Fig.\ref{fig:RAA_fit_with_each_mechanism}. We can see that by tuning the transport coefficient, collisional or radiative energy loss alone also provide reasonable description of the experimental data, though not as good as the case with both energy loss mechanisms included. To distinguish between different energy loss mechanisms, one may need explore more observables for heavy flavors.

In Fig.\ref{fig:corr_compare}, we investigate the angular correlation functions of $c\bar{c}$ pairs after they propagate through a QGP medium. We start with the back-to-back production of initial $c$ and $\bar{c}$ pairs and find that after they traverse the medium, the correlation function still peaks around $\pi$ if one only considers gluon radiation. On the contrary, a peak near 0 is observed if one only included the collisional energy loss. This indicates that unlike the back-to-back initial state, low energy $c\bar{c}$ pairs tend to move collinearly in the end because of boost by the collective flow of the medium.

In Fig.\ref{fig:ccbar-show}, we use an improved initialization procedure for $c\bar{c}$ production -- the Monte-Carto next-to-leading-order (MCNLO) production plus Herwig vacuum radiation \cite{Nahrgang:2013saa}. Similar to Fig.\ref{fig:corr_compare}, while the pure gluon radiation does not change the angular correlation function compared with the initial one, the pure collisional energy loss leads to a peak around 0. 
A more realistic analysis is implemented in Fig.\ref{fig:DDbar-show} where we loop each $D$ meson over all $\bar{D}$ within a collision event and analyze all possible $D\bar{D}$ pairs. We observe that the shapes of the correlation functions displayed in Fig.\ref{fig:DDbar-show} resemble those in Fig.\ref{fig:ccbar-show} except that a large background contributed from uncorrelated $D$ and $\bar{D}$ is observed. Although such correlation functions are dependent on different models, they may still provide a deeper insight into heavy quark energy loss mechanisms if a detailed comparison can be made in the future between theory and experiment.

\section{Summary}

We have studied heavy flavor evolution and hadronization in heavy-ion collisions. The quasi-elastic scatterings of heavy quarks off light partons and the medium-induced gluon radiation are simultaneously included in our modified Langevin framework. The subsequent hadronization process is calculated with a fragmentation plus recombination model. We find significant contribution from gluon radiation to heavy quark energy loss at high $p_\textnormal{\tiny T}$, and the inclusion of recombination process may enhance $D$ meson production at intermediate $p_\textnormal{\tiny T}$. Our calculations provide good descriptions of $D$ meson $R_\textnormal{\tiny AA}$ at the LHC. In addition, we explore the angular correlation functions of heavy flavor pairs and find them a potential candidate to distinguish different energy loss mechanisms of heavy quarks inside QGP.

\section*{Acknowledgments}

We are grateful to Prof. Berndt M\"uller and Dr. Marlene Nahrgang for helpful discussions. This work was supported by the U.S. Department of Energy Grant No. DE-FG02-05ER41367  and Natural Science Foundation of China (NSFC) under grant No. 11375072.





\bibliographystyle{elsarticle-num}
\bibliography{SCrefs}







\end{document}